\documentclass[preprint,12pt]{elsarticle}

\usepackage{amssymb}
\usepackage{amsmath}
\usepackage{graphicx}

\journal{Journal of Manufacturing Processes}

\begin{document}

\begin{frontmatter}



\title{High-Repetition-Rate Projection Multiphoton Lithography for Large-Area Sub-Micron 3D Printing}


\author{S. Papamakarios, M. Manousidaki, M. Stavrou, D. Gray, M. Farsari} 

\affiliation{organization={Institute of Electronic Structure and Laser, Foundation for Research and Technology Hellas (IESL-FORTH)},
            addressline={N. Plastira 100}, 
            city={Heraklion},
            postcode={71305}, 
            state={Crete},
            country={Greece}}

\begin{abstract}
High-resolution lithographic techniques are often limited by low volumetric throughput, since there is no universal and scalable manufacturing process that can produce 3D metasurfaces. In this work, we demonstrate a high-speed holographic 3D printing platform based on spatiotemporal beam shaping, exceeding the repetition rate while keeping the resolution high. The system integrates a femtosecond laser source with a spectral pulse compressor and a beam shaper to project uniform, axially confined light fields to project patterns directly on the advanced photoresists using a Digital Micromirror Device (DMD). We investigate the process window for rapid polymerization, optimizing the photoinitiator choice to eliminate thermal crosstalk at high repetition rates. Using this setup, we achieve a production throughput of $2.3*10^6$ voxels/s with sub-micron resolution ($<400$ nm). The system’s reliability is validated through the fabrication of large-area woodpile-like lattices and uniform micropillar arrays, establishing a workflow for scalable manufacturing of micro-optical components.
\end{abstract}

\begin{graphicalabstract}
\includegraphics[scale=0.49]{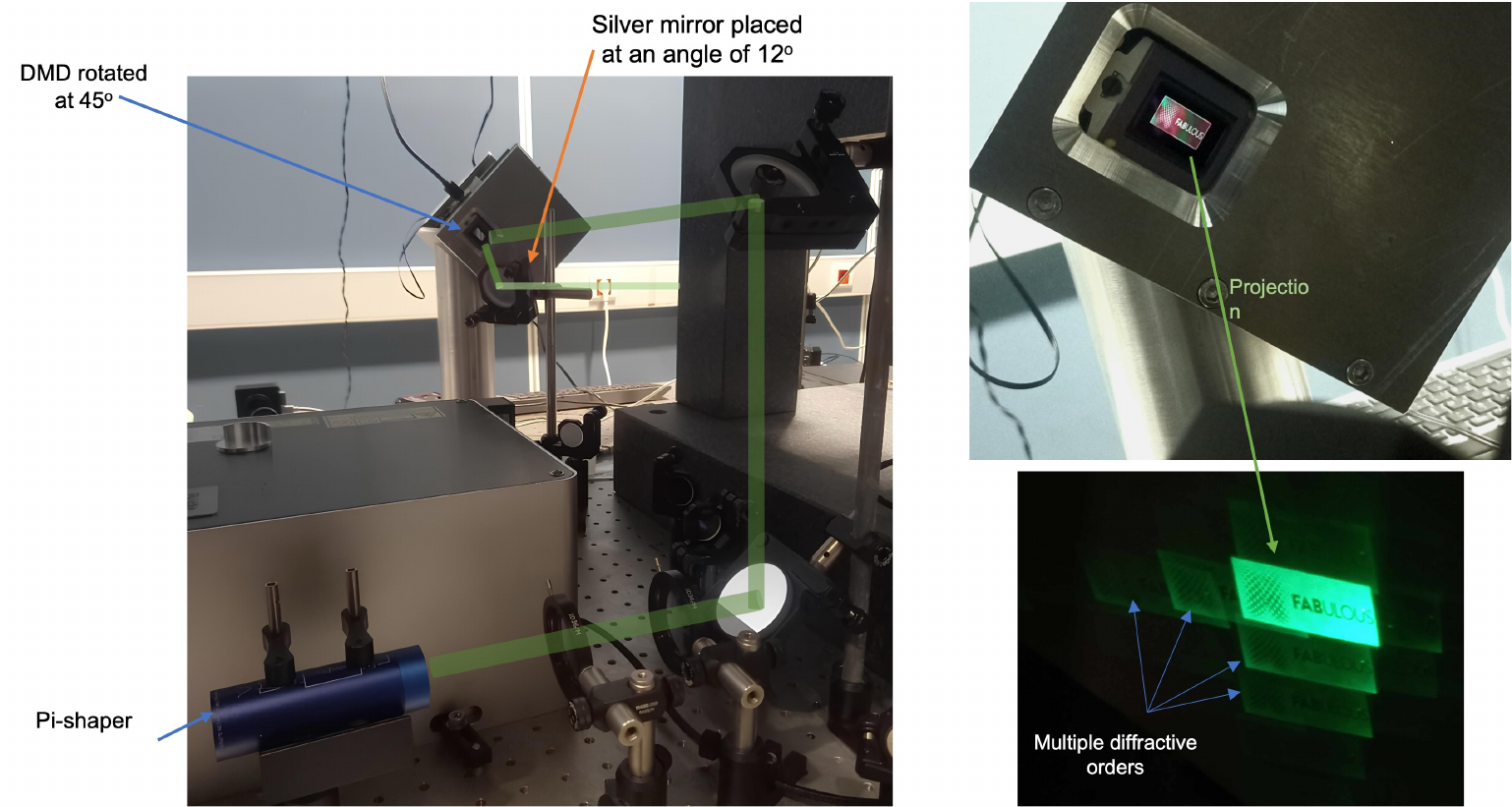}
\end{graphicalabstract}

\begin{highlights}
\item Exceeding $2.3*10^6$ voxels/s and $<60$ fs pulse duration with repetition rate 500 kHz
\item Fabrication of sub-400 nm resolution micropillar arrays
\item A holographic 3D printing platform using spatiotemporal beam shaping for large-area fabrication while keeping high resolution
\end{highlights}

\begin{keyword}
Multiphoton Lithography \sep Rapid fabrication


\end{keyword}

\end{frontmatter}



\section{Introduction}
\label{sec1}
High resolution additive manufacturing techniques are crucial for the fabrication of next-generation micro-optics~\cite{Liberale2010, Wang2023, Berglund2022}, metamaterials~\cite{Zyla2025AMT, Papamakarios2025} and scaffolds for tissue engineering~\cite{Flamourakis2020, Melissinaki2011}, where including a 3D geometry can give completely different capabilities compared to standard 2D elements~\cite{Katsantonis2023, Melissinaki2015_Sensor}. Currently Multiphoton Polymerization (MPL)~\cite{Kawata2001, Zyla2024, Malinauskas2015, Malinauskas2013, Lin2020}, a true mask-less 3D printing process relying on Direct Laser Writing (DLW), is the superb fabrication technique for sub-micron structures reaching resolutions even below 100 nm~\cite{skliutas2025multiphoton}. However, standard MPL suffers from proximity effects and out-of-plane hot spots~\cite{He2015}, since it is relying on point-by-point fabrication approach, where volumetric throughput is fundamentally limited by the system's components such as movement of the stages or resolution of galvanometric scanners. Keeping the resolution high while there is need for fabrication over a large area requires significant fabrication time for the devices~\cite{NatComm2023}, and lacks repeatability and scalability for potential industrial manufacturing applications where high-speed combined with high-resolution is essential. To overcome these issues, the research shifted towards projection lithography employing a DMD to polymerize simultaneously millions of voxels per second, keeping the resolution high, using a dedicated advanced novel optical system for MPL.

Despite the promise of projection lithography, scaling these systems to the sub-micron regime presents significant optical and thermal challenges. First, maintaining axial (z-axis) confinement without scanning is difficult; standard holographic beams often suffer from elongated focal volumes compared to tightly focused 2PP points~\cite{He2015, Zyla2024Light}. Second, and perhaps more critical for manufacturing, is thermal instability. Increasing the laser repetition rate to boost throughput often leads to local heat accumulation ("hot spots")~\cite{Stavrou2025, Ladika2025} and uncontrolled bulk polymerization. Furthermore, the Gaussian intensity profile typical of femtosecond lasers results in non-uniform energy distribution across the projection field, making it impossible to fabricate large arrays of identical structures with consistent dimensional accuracy~\cite{OptLaserTech2019, Yang2022}.

To develop the projection layer-by-layer 3D printing by spatial and temporal pulse shaping, holography of temporally and spatially shaped pulses will be used to generate 2D printed patterns~\cite{Saha2019}. This will allow simultaneous printing, reduce proximity effects and out-of-plane hot spots~\cite{He2015}. By separating the component wavelengths of a femtosecond laser pulse and then recombining them at a single plane, laser pulses become stretched in time, reducing the pulse intensity. As the pulse approaches the plot pattern plane, the pulse becomes shorter until it reaches the focus where all the component wavelengths recombine to form the shortest, highest intensity pulse~\cite{Somers2021}. After further propagation the wavelengths separate out again, reducing the intensity. Previous implementations of this technique were limited by the repetition rate of the laser used (in the kHz range) as well as a layer-by-layer approach with laser off time between each layer~\cite{OptLaserTech2019}.

To overcome these limitations, we implemented a projection system using a compact, high-power femtosecond fiber laser operating at MHz rates. This offers a major practical advantage over previous works that relied on expensive, bulky regenerative amplifiers~\cite{He2015, OptLaserTech2019}. Unlike amplifiers, which generate high-energy pulses that can damage the DMD (often forcing users to attenuate the beam and waste power), the MHz fiber laser distributes the energy over millions of pulses. This allows us to use the full average power of the source for rapid polymerization without burning the micro-mirrors, effectively solving the damage issue using simpler, more accessible hardware.

In this work, we present the design and process validation of a high-throughput projection 3D printing platform. The system incorporates a custom pulse compression module to reduce pulse duration to $<60$ fs and optimizes the photoinitiator chemistry (comparing BIS vs. Irgacure 369) to define a thermally stable process window. We validate the system’s performance by fabricating large-area micropillar arrays and woodpile lattices with sub-500 nm resolution. Under optimized conditions, the system demonstrates a stable volumetric throughput of $2.3*10^6$ voxels/s, offering a scalable pathway for the mass production of functional micro-structures.

\section{Experimental Setup}
\label{sec2}


\subsection{Prototype design and system layout}
The overall optical layout is schematically presented in Figure \ref{fig1} illustrating the sequence of components from the femtosecond laser source to the fabrication plane. This setup enables the generation of high repetition rates of polymerization process and rapid fabrication of arbitrary geometries and metasurfaces. 

\subsubsection{Irradiation source}
\begin{figure}[t]
\centering
\includegraphics[scale=0.5]{ 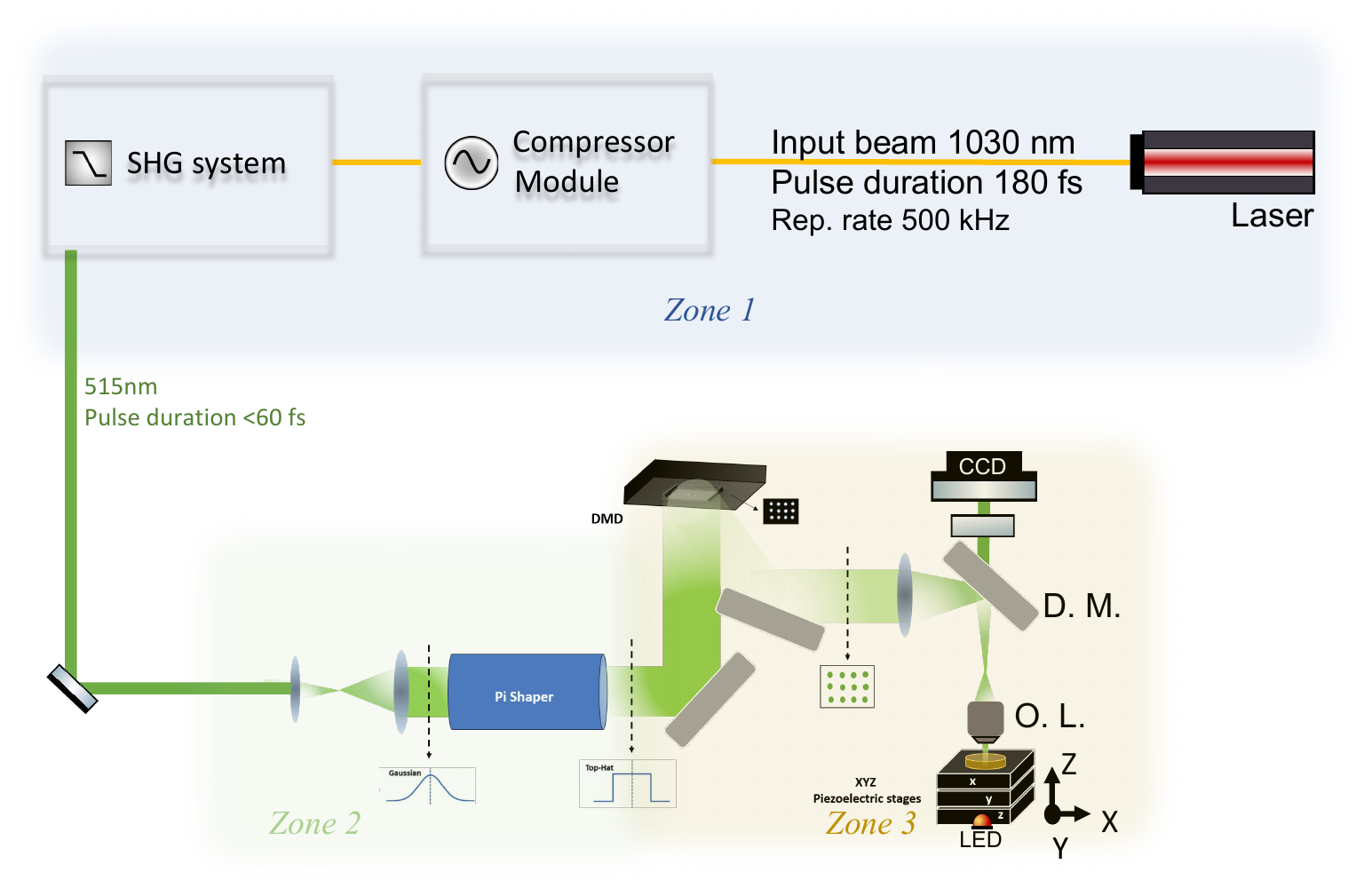}
\caption{Schematic of the integrated optical system showing the laser source, pulse compression module, SHG, and the spatiotemporal beam shaping path.}\label{fig1}
\end{figure}

The irradiation source of the setup is an Aeropulse FS10 from NKT Photonics, operating at 1030 nm, as well as at 515 nm, with a pulse duration of 180 fs and tunable repetition rate from 500 kHz up to 20 MHz, and maximum output power 10 W. This is the beginning of Zone 1 as it is indicated in Figure \ref{fig1}.

\subsubsection{Pulse stretcher}

We employ a pulse compressor (MIKS1S from n2 Photonics) to shorten the pulse duration by introducing spatial chirp (stretching the pulse in space). This spatiotemporal beam shaping results in a pulse duration of $<60$ fs, measured after the second harmonic generator (SHG) module. The system utilizes the IR output of the laser (1030 nm) operating at a 500 kHz repetition rate. This repetition rate was specifically selected to balance volumetric throughput with thermal stability; it provides a sufficient time interval between pulses in the focal volume to reduce proximity effects and eliminate out-of-plane hot spots, while simultaneously maintaining high peak intensity for rapid fabrication. Additionally, the 500 kHz rate was chosen to operate safely within the damage thresholds of the DMD, considering the device's micro-mirrors possess a default refresh rate in the kHz regime. Following compression, an SHG setup is used to produce a stable, compressed pulse with a central wavelength of 515 nm, marking the end of Zone 1 (Figure \ref{fig1}).

\begin{figure}[h]
\centering
\includegraphics[scale=0.4]{ 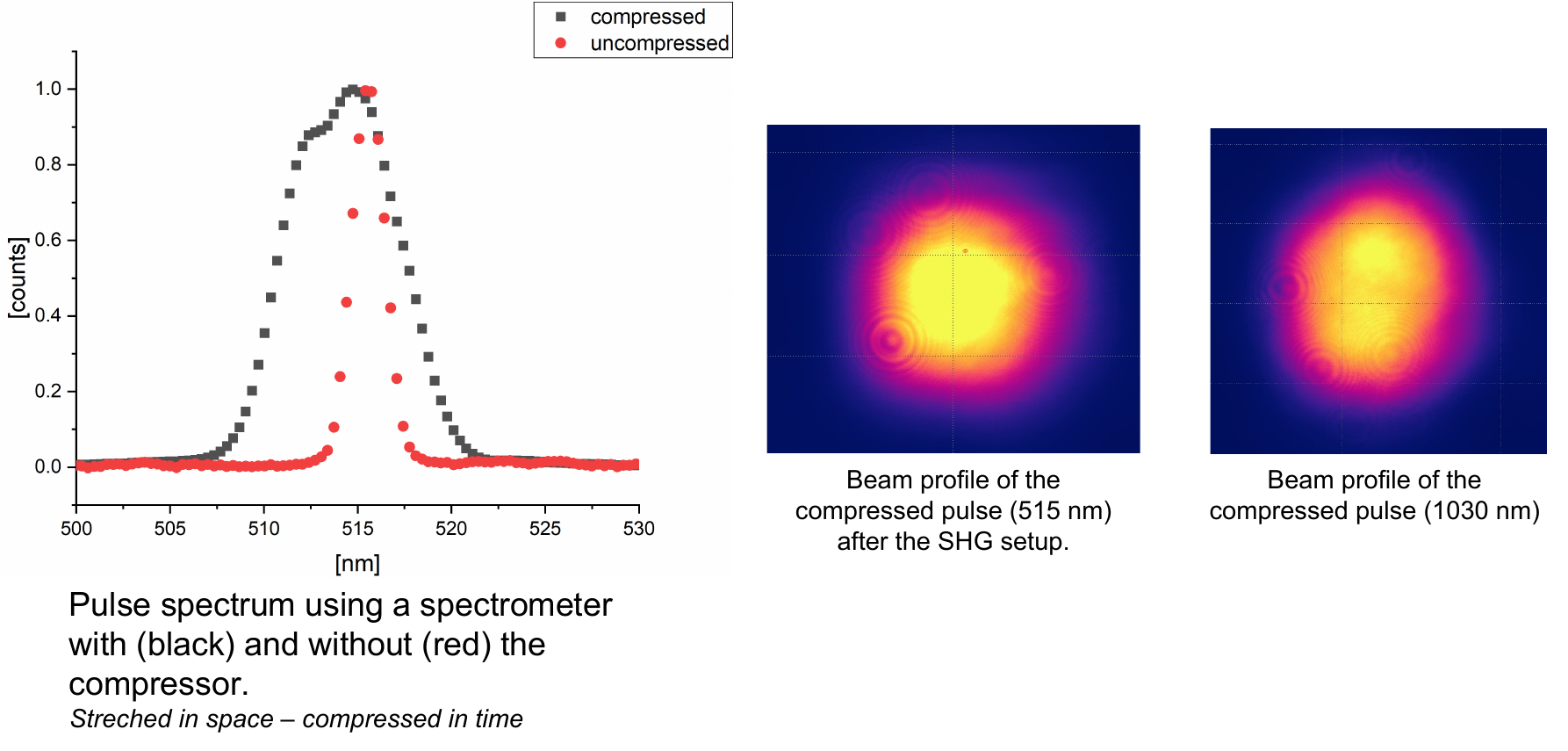}
\caption{Comparison of spectral analysis before and after the compressor module (left), and quality of the profile of the laser beam accordingly (right).}\label{fig2}
\end{figure}

The performance of the pulse stretcher was quantified using a spectrometer to analyze the spatial spectrum of the laser beam before and after the module. As depicted in Figure \ref{fig2}, we observe a spatial spectrum stretched by a factor of $>3$, which corresponds to the temporal compression of the initial 180 fs pulse down to $<60$ fs. The quality of the laser beam was also characterized using a beam-profile camera, confirming that the Gaussian shape and beam quality are preserved after the module, as shown in Figure \ref{fig2}.

\subsubsection{Top-hat beam formation}
\begin{figure}[h]
\centering
\includegraphics[scale=0.45]{ 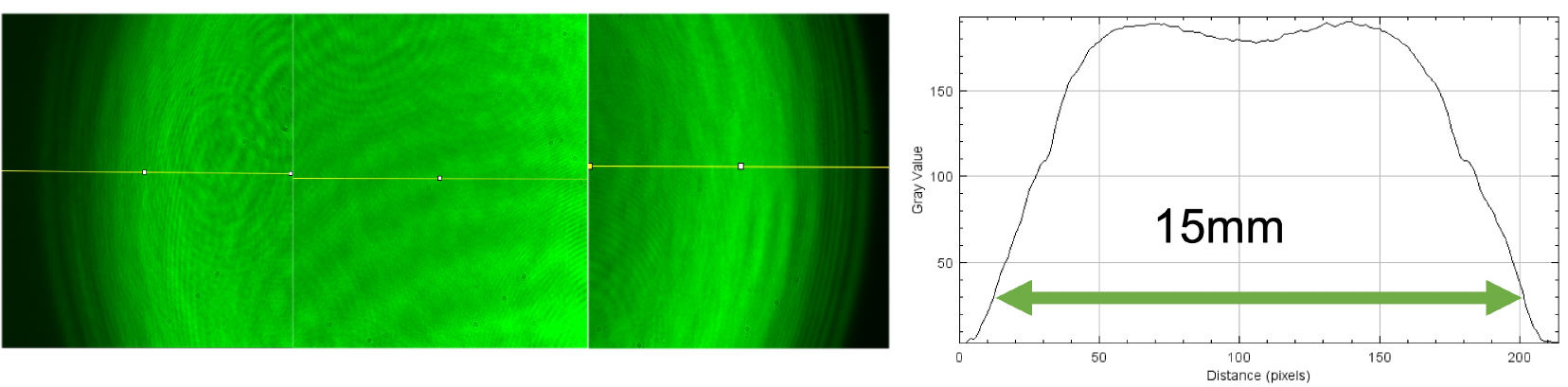}
\caption{Top-hat beam profile after the $\pi$-shaper (left), and normalized intensity highlighting the output diameter of the beam and the uniformity of the intensity (right).  }\label{fig3}
\end{figure}

To ensure intensity uniformity of the projection from the DMD, a $\pi$-shaper was used to convert the Gaussian beam at the output of the SHG module (Figure \ref{fig2}, right) into a top-hat beam, ensuring the pulse intensity is consistent across the entire beam diameter. Using a Keplerian telescope, we expanded the beam to 15 mm (Figure \ref{fig3}) to match the input pupil of the $\pi$-shaper and ensure homogeneous projection from the DMD. This expansion results in a collimated flat-top output beam with a diameter of 15 mm, designed to slightly overfill the active area of the DMD ($14.5 \text{ mm} \times 8.2 \text{ mm}$). Consequently, the mirrors at the corners of the array receive the same incident intensity as those in the center. This uniformity is critical for polymerizing the entire field of view simultaneously without intensity gradients, thereby avoiding hot spots or underexposed regions within the focal volume. The formation of the top-hat beam profile is highlighted in Figure \ref{fig1}, Zone 2.

\subsubsection{Diffractive projection from Digital Micromirror Device and spatiotemporal beam shaping}

The DLP6500 Digital Micromirror Device (DMD) was utilized as a programmable diffractive optical element (DOE). Given the periodic microstructure of the mirror array ($7.6~\mu$m pitch), the device induces significant angular dispersion upon the broadband femtosecond spectrum. To optimize the photon transport efficiency, the DMD chassis was rotated azimuthally by $45^{\circ}$ relative to the optical table. This alignment orients the orthogonal micromirror tilt axes ($\pm 12^{\circ}$) perpendicular to the plane of incidence, thereby satisfying the blazed grating condition. Consequently, the specular reflection from the ``ON'' state mirrors is coupled efficiently into the fundamental diffraction order, minimizing energy loss to parasitic higher-order modes. Maximum repetition refresh rate of the DMD is by default 10 kHz, and the total screen size and available pixels are 1920 $\times$ 1080 (approximately 2 million micromirrors) covering a total area of $13.5$ mm $\times$ $8.2$ mm.

\subsubsection{Simultaneous Spatiotemporal Focusing (SSTF)}
\begin{figure}[h]
\centering
\includegraphics[scale=0.45]{ 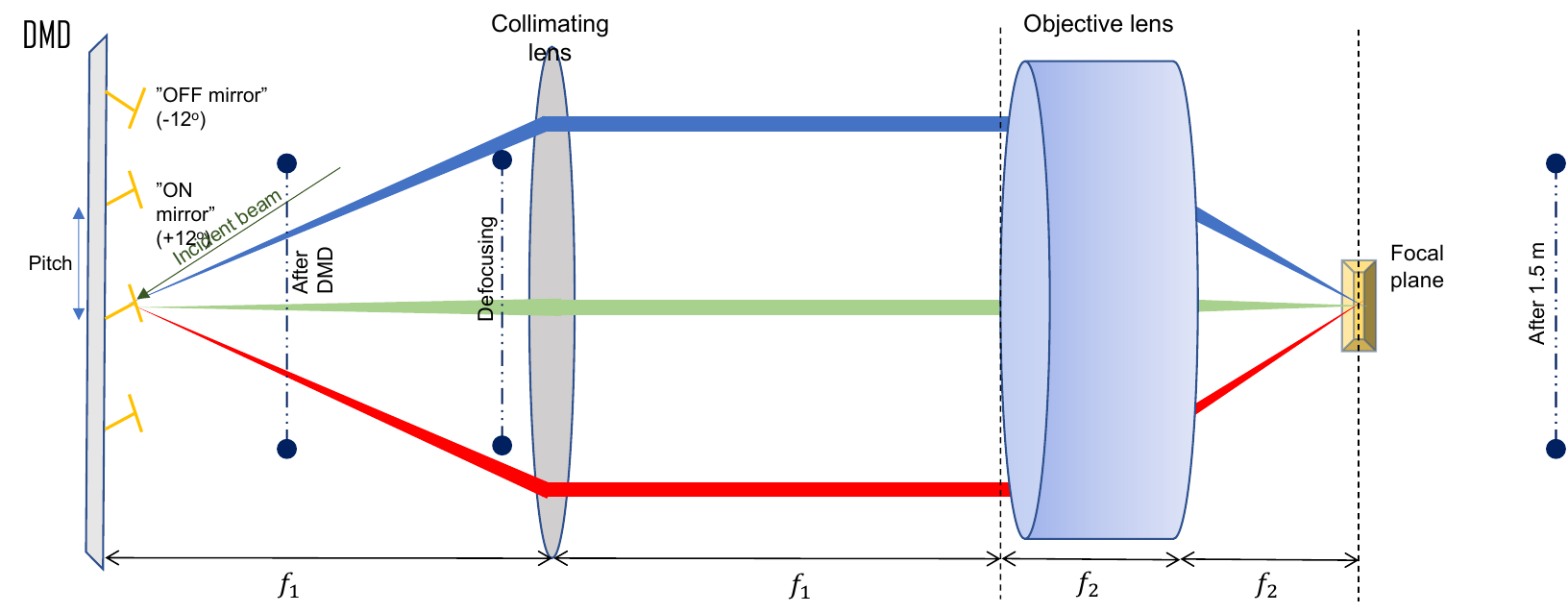}
\caption{Schematic presentation of the 4f configuration system after the DMD, where the temporal focusing is highlighted.}\label{fig4}
\end{figure}

The system exploits the angular dispersion introduced by the DMD to achieve axial confinement via Simultaneous Spatiotemporal Focusing (SSTF). The diffracted beam, characterized by a spatial chirp, is relayed through a 4f optical system. The high-numerical-aperture objective (NA 1.4) functions as a Fourier transform lens, performing the recombination of the spectral components at the focal plane, as it is depicted in Figure \ref{fig4}.

In the geometric focal volume, the spatially dispersed frequencies overlap, and the group delay dispersion (GDD) accumulated through the optical train is compensated, resulting in the isochronous superposition of the spectral components. Consequently, as the pulse propagates towards the focal plane, it undergoes temporal compression, reaching its maximum peak intensity exactly at the focus where the pulse duration is restored to its bandwidth-limited minimum ($<60$ fs). Outside this region, the pulse exhibits coupled spatial and temporal chirp, resulting in a rapid decrease in peak intensity. This mechanism ensures that the two-photon polymerization threshold is exceeded only within a confined Rayleigh range, effectively suppressing out-of-focus background polymerization.

The schematic in Figure \ref{fig4} represents the last part of the setup, named Zone 3. The fabrication process is controlled using a dedicated advanced software "CICADA". The sample is placed on high precision linear stages (L-731 Precision XY Stage, V-551 Precision Linear Stage) which allow the rapid fabrication process with high accuracy in stitching, offering large area fabrication up to 20 cm. 
\section{Materials and methods}
\label{sec3}


\subsection{Impact of photoinitiator}

We investigated the polymerization stability of two formulations based on the SZ2080~\cite{Ovsianikov2008} hybrid resin. Initial trials utilized 4,4$'$-bis(diethylamino) benzophenone (BIS). While highly reactive, BIS exhibits a non-negligible absorption peak near 515 nm~\cite{Ladika2025}, which results in large volume of hot spots and polymerization defects. Under 500 kHz irradiation, this residual linear absorption induces a parasitic thermal background. The resultant exothermic accumulation creates a positive feedback loop, lowering the local polymerization threshold and causing uncontrolled radical diffusion. This was empirically observed as thermal blooming and solvent vaporization (boiling) within the focal volume~\cite{Stavrou2025}.

\begin{figure}[h]
\centering
\includegraphics[scale=0.53]{ 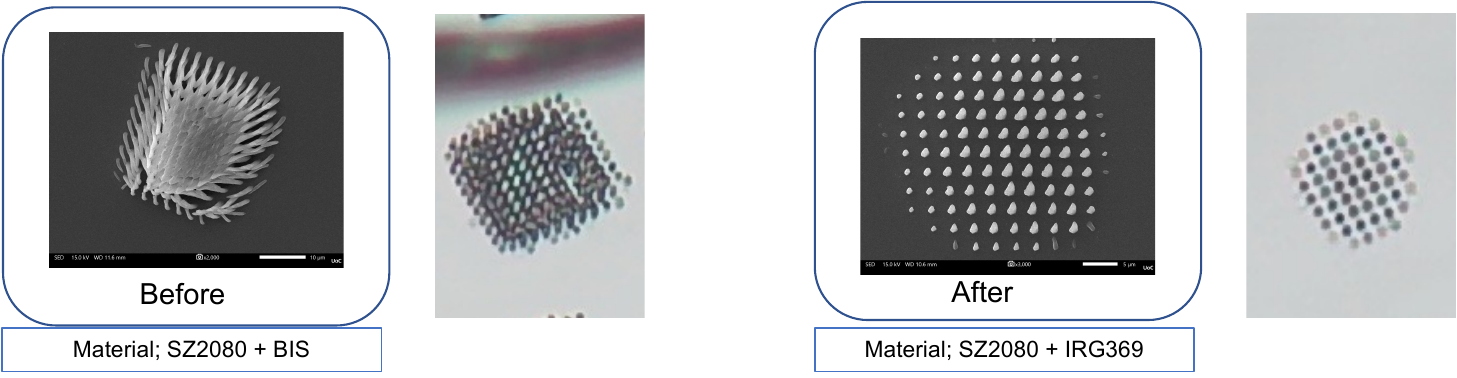}
\caption{SEM and microscope images of micropillars arrays fabricated employing DMD and spatiotemporal beam shaping, showing the impact of the photoinitiator in the fabrication process with BIS (left) and IRG369 (right).}\label{fig5}
\end{figure}

Conversely, Irgacure 369 possesses a negligible absorption peak in the green spectral window, ensuring that radical generation is driven exclusively by non-linear two-photon absorption. This separates the polymerization mechanism from bulk thermal loading. Experimental validation confirmed that the Irgacure-sensitized resin remained thermally stable at 500 kHz, producing sharply defined voxels with no evidence of proximity effects as highlighted in Figure \ref{fig5}, consistent with the high degree of conversion expected for zirconium-silicon hybrids.

\section{Results and Discussion}

\subsection{Fabrication of micropillars}
To validate the spatial resolution limit imposed by the SSTF optics, high-aspect-ratio micropillar arrays were fabricated. These structures serve as a physical representation of the system’s Point Spread Function (PSF).

Under optimized exposure conditions (200--400 ms), we achieved a minimum single spot feature size of 352 nm. The pillars exhibited high verticality with uniform cross-sections throughout the structural height ($<3~\mu$m). This result confirms that the bandwidth-limited pulses ($<60$ fs) effectively drive the polymerization non-linearity, confining the reaction strictly to the focal volume defined by the SSTF light sheet, effectively suppressing the axial elongation typical of standard projection beams. Also, the high aspect ratio and the stability of the micropillars indicates the high quality of the simultaneous polymerization at the Fourier plane. 

Different pillars in diameter and height were produced using specific binary images were white indicates the mirrors that are in "ON" state, and black the "OFF" state. Changing the number of pixels that are "ON" different pillars were processed with high repeatability and stability, controlling accurately the diameter and the height of the structures.

\begin{figure}[h]
\centering
\includegraphics[scale=0.51]{ 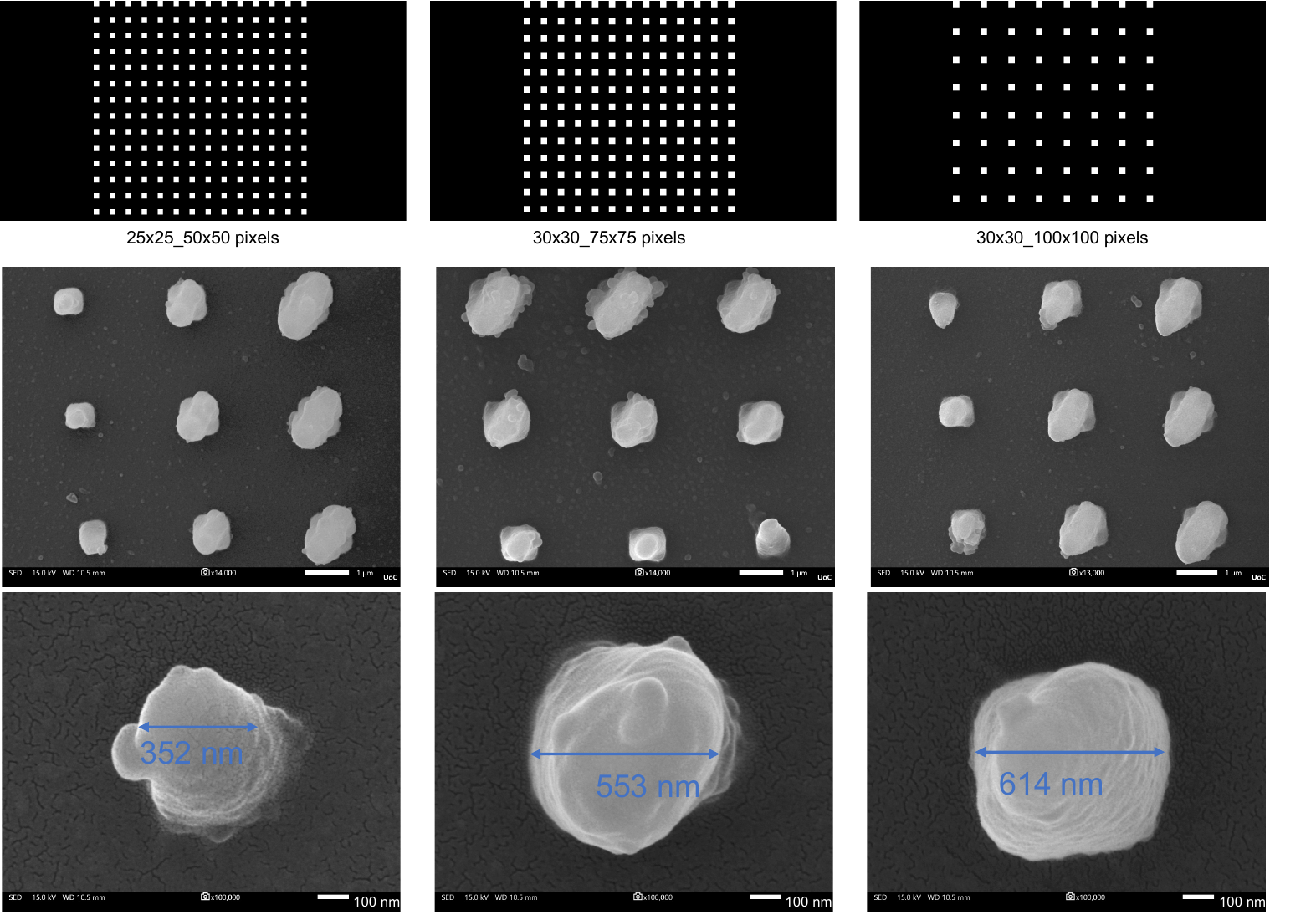}
\caption{Presentation of masks that were used for the fabrication of micropillars arrays with varying diameter and height using different number of "ON" and "OFF" pixels, and, SEM images of the processed pillars highlighting the repeatability and resolution of the process.}\label{fig6}
\end{figure}

\subsection{Large-Scale Volumetric Fabrication}
\begin{figure}[h]
\centering
\includegraphics[scale=0.5]{ 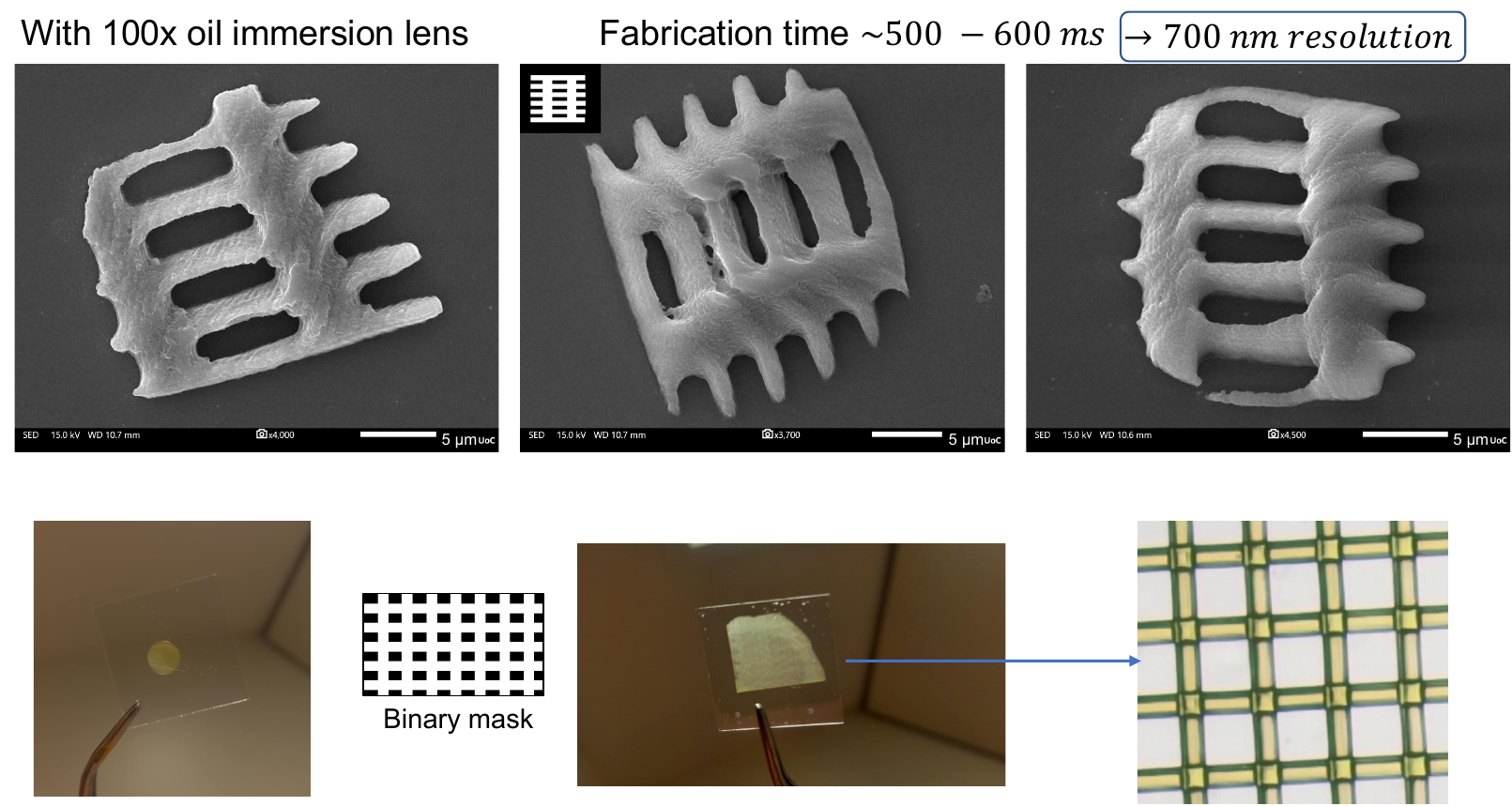}
\caption{SEM images of grids fabricated using  projection 3D printing by spatiotemporal ultrashort pulses and large area fabrication using stitching process and high accuracy linear stages.}\label{fig7}
\end{figure}

First, to evaluate structural continuity, woodpile lattice structures were fabricated via step-and-repeat projection. The resulting lattices (Figure \ref{fig7}), which were processed using 100$\times$ oil immersion objective lens, exhibit seamless interfacial bonding, confirming that the Irgacure 369 formulation prevents the voxel expansion that typically leads to over-polymerization at field boundaries, reaching to a resolution of 700 nm without hot spots and unregulated polymerization near the "ON" pixels on the image that was used for projection.

Subsequently, to demonstrate the high-speed capabilities over macroscopic areas, the system was configured with a 40$\times$ oil immersion objective lens. In this configuration, the increased field of view allowed for the rapid patterning of a $1.2 \text{ cm} \times 1.2 \text{ cm}$ area in just 2 minutes, using stitching process and the capabilities of the high travel velocity of the ultra-precision linear translation stages. This corresponds to a stable volumetric throughput of $2.3 \times 10^6$ voxels/s, effectively balancing industrial-grade speed with the resolution required for large-scale metasurface applications.

\subsection{Gray-Scale Lithography (Micro-Optics)}
Finally, we exploited the dependence of the polymer’s degree of conversion on accumulated fluence to fabricate analog micro-optical elements. By spatially modulating the exposure duration of circular masks ($50-200$ ms), we successfully tuned the curvature and sag height of hemispherical micro-lenses (diameters $10-33~\mu$m). The optical-grade smoothness of the lens surfaces indicates a deterministic polymerization regime, free from the stochastic roughness associated with thermally driven processes.

The fabrication of refractive micro-optical elements, such as microlens arrays, presents a unique challenge for standard point-scanning Two-Photon Polymerization (TPP). Because conventional systems trace the structure voxel-by-voxel, the fabrication time scales linearly with the volume, rendering large-area patterning prohibitively slow. More critically, the serial nature of the scanning process inherently induces periodic surface modulations—known as ``hatching artifacts''—corresponding to the step size of the galvanometric mirrors. These artifacts act as parasitic diffraction gratings that scatter light and degrade the imaging quality of the lens, often requiring time-consuming post-processing or extreme over-sampling to mitigate.

In contrast, the SSTF projection platform fundamentally decouples fabrication time from structural volume. By projecting a continuous optical field, the entire effective aperture of the microlens is defined simultaneously, eliminating the discrete scan lines responsible for hatching ripples. We demonstrated this by generating hemispherical lenses with diameters ranging from 10 to 33~$\mu$m through precise fluence control ($50$--$200$ ms). This approach not only increases throughput by several orders of magnitude compared to scanning but also yields a geometric continuity essential for high-performance imaging optics. This capability complements our recent demonstrations of electromagnetic metasurfaces \cite{Zyla2025AMT, Papamakarios2025} by extending the platform’s utility into the regime of visible-light refractive components.

\begin{figure}[h]
\centering
\includegraphics[scale=0.55]{ 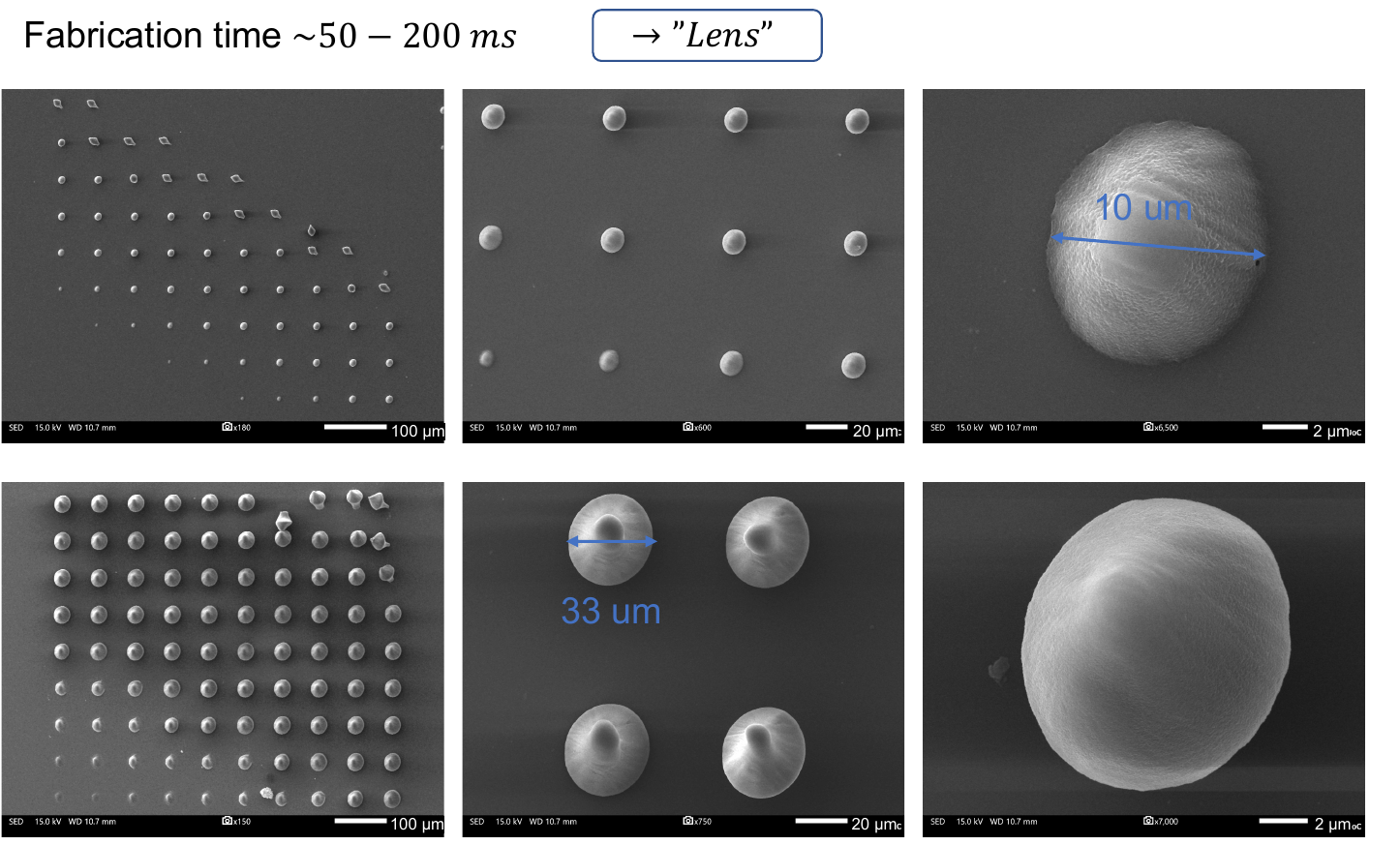}
\caption{SEM images of microlenses arrays fabricated using different number of pixels in the focal spot. The diameter and the curvature is controlled depending on the number of pixels that are "ON" and "OFF".}\label{fig8}
\end{figure}

\section{Conclusion}

In this work, we have demonstrated a compact and cost-efficient holographic 3D printing platform that effectively bridges the gap between standard point-scanning TPP and complex, high-end projection systems. By integrating a commercial Digital Micromirror Device (DMD) with a custom-built dispersion compensation module, we successfully delivered bandwidth-limited pulses of \textbf{$<$60 fs} to the sample plane. This temporal compression was critical for maximizing the two-photon excitation efficiency, enabling a stable fabrication process at a repetition rate of \textbf{500 kHz}.

The system’s performance was validated through the fabrication of dense, high-aspect-ratio micropillar arrays, achieving a lateral resolution of \textbf{$<$400 nm} and verifying the optical sectioning capability of the SSTF scheme. Furthermore, we demonstrated the scalability of the platform by stitching large-area woodpile lattices at a continuous throughput of \textbf{$2.3 \times 10^6$ voxels/s}. These results confirm that by optimizing the spatiotemporal pulse profile, high-speed volumetric manufacturing can be achieved without the need for high-power amplified laser sources, offering a robust and accessible solution for rapid micro-prototyping.

\section{Acknowledgments}
The authors acknowledge funding by European Union through project FABulous (HORIZON-CL4-2022-TWIN-TRANSITION- 01-02, GA:101091644). The research project was co-funded by the Stavros Niarchos Foundation (SNF) and the Hellenic Foundation for Research and Innovation (H.F.R.I.) under the 5th Call of “Science and Society” Action – “Always Strive for Excellence – Theodore Papazoglou” (Project Number: 9578.).The authors acknowledge Ms Aleka Manousaki for the excellent SEM support.

\newpage
\appendix
\section{Pictures of the fabrication setup}
\begin{figure}[h]
\centering
\includegraphics[scale=0.33]{ 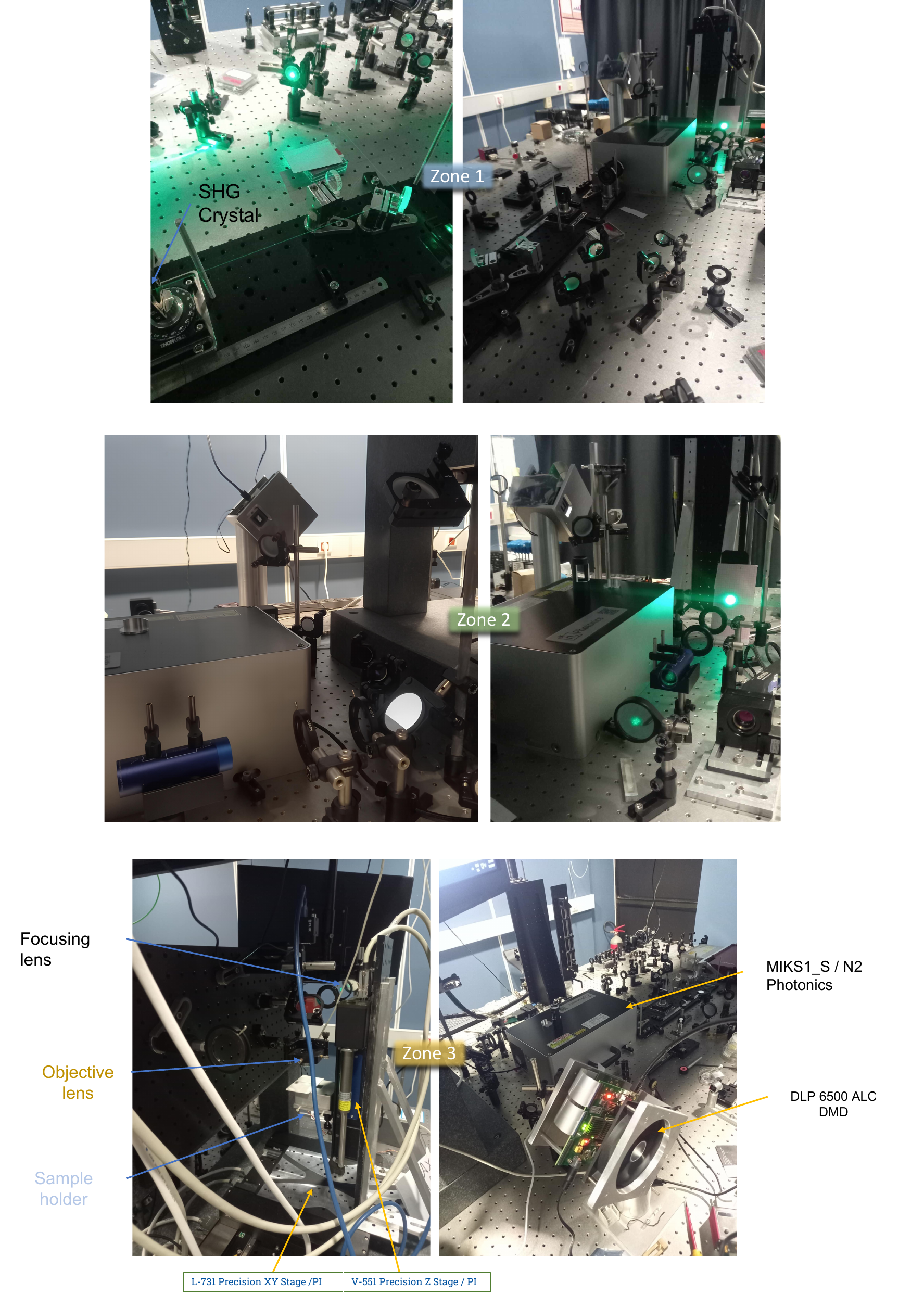}
\caption{Pictures of the complete setup indicating the 3 different zones.}\label{fig9}
\end{figure}

\newpage

\bibliographystyle{unsrt}
\bibliography{reference}

\end{document}